\documentclass[a4paper, 10pt]{report}

\usepackage[cp1250]{inputenc}
\usepackage[T1]{fontenc}
\usepackage{amsfonts}
\usepackage{graphicx}
\usepackage{amsmath}
\usepackage{caption}

\usepackage{amssymb}

\usepackage{fancyhdr}
\usepackage{bibtopic}
\usepackage{color}
\usepackage{ulem}
\usepackage[toc,page]{appendix}
\usepackage{setspace}
\usepackage{cite}

\usepackage{soul} 

\usepackage{xcolor}

\AtBeginDocument{}
\usepackage{etoolbox}
\patchcmd{\thebibliography}{\chapter*}{\section*}{}{}

\makeatletter
\renewcommand{\thesection}{%
  \ifnum\c@chapter<1 \@arabic\c@section
  \else \thechapter.\@arabic\c@section
  \fi
}
\makeatother

\patchcmd{\tableofcontents}{\chapter*}{\section*}{}{}

\numberwithin{equation}{section} 

\let\OLDthebibliography\thebibliography
\renewcommand\thebibliography[1]{
  \OLDthebibliography{#1}
  \setlength{\parskip}{0pt}
  \setlength{\itemsep}{3.5pt plus 1ex}
}

\usepackage[linktocpage]{hyperref}
\definecolor{darkred}{rgb}{0.5,0,0}
\definecolor{darkpurple}{rgb}{0.5,0,0.5}
\definecolor{darkblue}{rgb}{0,0,0.5}
\hypersetup{ colorlinks,
linkcolor=darkblue,
filecolor=darkpurple,
urlcolor=darkred,
citecolor=darkpurple }

\usepackage[symbol]{footmisc}
\usepackage{cancel}

\begin{document}

\allowdisplaybreaks
\setlength{\abovedisplayskip}{3.5pt}
\setlength{\belowdisplayskip}{3.5pt}
\abovedisplayshortskip
\belowdisplayshortskip

{\setstretch{1.0}

{\LARGE \bf \centerline{
Interaction of fluids described through relative motion
}}
{\LARGE \bf \centerline{
and application to Bianchi type-I spacetimes
}}

\vskip 1cm
\begin{center}
{Peter M\'esz\'aros\footnote[1]{peter.meszaros@fmph.uniba.sk}, Daniel Ra\v{c}ko\footnote[2]{daniel.racko@fmph.uniba.sk}}

\vskip 2mm {\it Department of Theoretical Physics, Comenius
University, Bratislava, Slovakia}

\vskip 2mm \today 
\end{center}

\section*{Abstract}

We study a model of two fluids with an interaction that depends on their relative motion. The interaction is described using the particle number density of each fluid measured in the rest frame of the other, providing a geometric interpretation of a class of interactions studied in relativistic multifluid theory. We derive the corresponding stress-energy tensor and show that its structure agrees with established results. In particular, the interaction introduces a dependence on the scalar product of the $4$-velocities, which can be expressed in terms of the relative speed of the fluid elements. Furthermore, the relative motion produces unequal pressures along and perpendicular to the direction of motion, together with an energy flux. As an application, we consider a Bianchi type-I universe dominated by one fluid, with the second fluid acting as a source of small anisotropy. We examine how the interaction influences the relative speed and the evolution of cosmological anisotropy. We identify conditions under which the anisotropy decays in the late-time limit, generalizing results for non-interacting fluids from the literature.

\vskip 3mm \hspace{4mm}
\begin{minipage}[t]{0.8\textwidth}
\noindent\rule{12cm}{0.4pt}
\vspace{-9mm}
\tableofcontents
\noindent\rule{12cm}{0.4pt}
\end{minipage}
\vskip 6mm

}

\section{Introduction}\label{sec:1}

Perfect fluid provides an adequate description of relativistic matter in a wide range of physical scenarios. It is fully characterized by its energy density $\rho$ and pressure $p$, measured in the rest frames of its volume elements, and by the $4$-velocity $u^{\mu}$ of these elements, usually supplemented by an equation of state $p(\rho)$. The simplest type of perfect fluid is one with a constant pressure-to-energy density ratio $w=p/\rho$. For radiation, it is $w=1/3$, and for ultrarelativistic matter, it is close to this value. This can be applied to the expansion of the early radiation-dominated Universe or to the interiors of compact objects such as neutron stars; however, the description of realistic neutron stars departs from the simplifying assumption of $w$ being constant\cite{Glen}. Perfect fluids with other constant values of $w$ can be used to describe the matter content of the Universe during later stages of its expansion, when the total energy density is dominated by the dark matter, $w=0$, and the dark energy, $w=-1$.

\vskip 2mm
A single scalar field can mimic a perfect fluid\cite{Madsen,Faraoni1,Faraoni2} in regimes that are applicable to both cosmology\cite{Ratra,Ellis} and hypothetical self-gravitating objects such as boson stars\cite{Kaup,Ruffini} or oscillatons\cite{Seidel}. This approach enables the derivation of more exotic equations of state directly from the matter Lagrangian, yielding models such as quintessence, $k$-essence\cite{Chiba,Picon1,Picon2}, or the Chaplygin gas\cite{Kamen,Bilic,Bento,Gorini}. In general, a perfect fluid cannot be fully described by a single scalar field, and its matter Lagrangian has to be constructed in a different way, as has been done in several works\cite{Taub,Schutz,Brown,Carter,Elze,Pop}. The perfect fluid Lagrangian is described through the field theory approach, where the role of fields is played by Lagrangian coordinates of the volume elements, and Lagrange multipliers are associated with thermodynamic quantities. In this paper, we will denote the Lagrangian coordinates as $\mathcal{X}^A$, associated with capital Latin indices $A=1,2,3$, and we will refer to the fictitious three-dimensional space parametrized by the Lagrangian coordinates as the material space. This approach has also made it possible to introduce interactions of a fluid with another fluid\cite{Andersson,Samuelsson,Jimenez1,Jimenez2} or with scalar field\cite{Ios,Ashi}, and to study non-minimal couplings to gravity\cite{Bertolami,Haghani,Boehmer}.

\vskip 2mm
The aforementioned interactions of a fluid with other matter fields or with gravity are usually described in terms of Lagrange multipliers\cite{Ios}, algebraic couplings\cite{Ashi}, or scalar products of particle fluxes\cite{Andersson,Samuelsson} or $4$-velocities\cite{Jimenez1,Jimenez2} of different fluids. In this work, we present an alternative approach based on a modification of the body metric $N^{AB}$, often used in the context of relativistic elasticity\cite{Carter2,Being}. It is defined as a push-forward of the inverse metric $g^{\mu\nu}$ with respect to a map from the spacetime to the material space, $x^{\mu} \to \mathcal{X}^A$,
\begin{eqnarray}\label{eq:def0}
N^{AB} = g^{\mu\nu} \mathcal{X}^A_{\phantom{A},\mu} \mathcal{X}^B_{\phantom{B},\nu},
\end{eqnarray}
where we use the $(-,+,+,+)$ signature of the spacetime metric, so that $N^{AB}$ is positive definite. The body metric measures the particle number density of the fluid as $n\propto\sqrt{\textrm{det}N}$, and the matter Lagrangian of a perfect fluid then depends on it. This density is measured in the rest frames of the fluid's volume elements. We will study a pair of fluids whose mutual interaction is formulated in terms of quantities that describe the particle number density of one fluid as measured in the reference frame of the volume elements of the other fluid, as opposed to the conventional density measured in the fluid's own reference frame. This approach reveals a new geometric interpretation of the already known form of the interaction, defined in terms of scalar products of the fluxes or $4$-velocities, and leads to the form of the stress-energy tensor found in the literature\cite{Andersson,Samuelsson,Jimenez1,Jimenez2}.

\vskip 2mm
In cosmology, two related classes of problems have been studied. The first addresses the anisotropic expansion of the Universe at the background level, where the anisotropy arises from the relative motion of non-interacting fluids\cite{King,Sandin,Harko}. The second examines interactions between fluids and their influence on cosmological perturbations on a homogeneous and isotropic background\cite{Jimenez1,Jimenez2}. Our work bridges these two approaches by investigating the effect of the fluid interaction on the anisotropic background expansion. We will describe the anisotropic expansion of the Universe by means of the Bianchi type-I metric, where we will focus on the case of small anisotropy.

\vskip 2mm
In addition to the already mentioned conventions, we will denote the Newtonian gravitational constant as $\kappa$, and use units in which the speed of light is set to unity, $c=1$.

\section{Stress-energy tensor}\label{sec:2}

Since the spacetime is four-dimensional and the material space is three-dimensional, by performing the pull-back of the body metric (\ref{eq:def0}) to the spacetime, we do not retrieve the full inverse metric $g^{\mu\nu}$, but rather its projection onto hypersurfaces orthogonal to the worldlines of the volume elements of the fluid $g^{\mu\nu}+u^{\mu}u^{\nu}$\cite{Carter2}. These are the local rest spaces of the volume elements. Therefore,
\begin{eqnarray}\label{eq:def}
N^{AB} = \left( g^{\mu\nu}+u^{\mu}u^{\nu} \right) \mathcal{X}^A_{\phantom{A},\mu} \mathcal{X}^B_{\phantom{B},\nu},
\end{eqnarray}
serves as an equivalent definition of the body metric (\ref{eq:def0}). It is then apparent that the body metric measures volumes from the point of view of reference frames that are at rest with respect to the volume elements of the fluid.

\vskip 2mm
Consider two fluids with two separate sets of Lagrangian coordinates $\mathcal{X}_1^A$ and $\mathcal{X}_2^A$ and two $4$-velocities $u_1^{\mu}$ and $u_2^{\mu}$. Hereafter, subscripts $1$ and $2$ label the first and second fluid. In addition to their body metrics,
\begin{eqnarray}\label{eq:defn}
N_1^{AB} = \left( g^{\mu\nu} + u_1^{\mu} u_1^{\nu} \right) \mathcal{X}^A_{1\phantom{|},\mu} \mathcal{X}^B_{1\phantom{|},\nu}, \quad N_2^{AB} = \left( g^{\mu\nu} + u_2^{\mu} u_2^{\nu} \right) \mathcal{X}^A_{2\phantom{|},\mu} \mathcal{X}^B_{2\phantom{|},\nu},
\end{eqnarray}
we may define also body metric-like objects that measure volumes of elements of one fluid from the point of view of the other fluid,
\begin{eqnarray}\label{eq:defa}
A_1^{AB} = \left( g^{\mu\nu} + u_2^{\mu} u_2^{\nu} \right) \mathcal{X}^A_{1\phantom{|},\mu} \mathcal{X}^B_{1\phantom{|},\nu}, \quad A_2^{AB} = \left( g^{\mu\nu} + u_1^{\mu} u_1^{\nu} \right) \mathcal{X}^A_{2\phantom{|},\mu} \mathcal{X}^B_{2\phantom{|},\nu}.
\end{eqnarray}
The matter Lagrangian of a pair of fluids interacting through the particle number densities of one fluid measured from the point of view of the other fluid then depends on the following quantities,
\begin{eqnarray}\label{eq:quantdef}
n_1 = n_0 \sqrt{\textrm{det}N_1}, \quad n_2 = n_0 \sqrt{\textrm{det}N_2}, \quad a_1 = n_0 \sqrt{\textrm{det}A_1}, \quad a_2 = n_0 \sqrt{\textrm{det}A_2},
\end{eqnarray}
where $n_0$ is a constant reference particle number density.
From now on, the subscript $(1,2)$ will be used as a shorthand for both fluids in such a way that $(2,1)$ will indicate fluid other than the fluid denoted by $(1,2)$.

\vskip 2mm

Since quantities defined in (\ref{eq:defn}) can also be defined as (\ref{eq:def0}), the quantities (\ref{eq:defa}) can be written in the form
\begin{eqnarray}
A_{(1,2)}^{AB} = N_{(1,2)}^{AB} + u^{\mu}_{(2,1)} \mathcal{X}^A_{(1,2),\mu} u^{\nu}_{(2,1)} \mathcal{X}^B_{(1,2),\nu},
\end{eqnarray}
which allows us to employ the matrix determinant lemma,
\begin{eqnarray}
\textrm{det} A_{(1,2)} & = & \left( \textrm{det} N_{(1,2)} \right) \left( 1 + N_{(1,2)AB} u^{\mu}_{(2,1)} \mathcal{X}^A_{(1,2),\mu} u^{\nu}_{(2,1)} \mathcal{X}^B_{(1,2),\nu} \right) \\ & = & \left( \textrm{det} N_{(1,2)} \right) \left( u_1 \cdot u_2 \right)^2, \nonumber
\end{eqnarray}
where $u_1\cdot u_2$ denotes the scalar product of the two $4$-velocities, $u_1\cdot u_2=g_{\mu\nu}u_1^{\mu}u_2^{\nu}$. The last equality is a consequence of projections to hypersurfaces orthogonal to worldlines of fluids that occur in the process of pulling back the body metric $N_{(1,2)AB}$ to the spacetime,
\begin{eqnarray}
N_{(1,2)AB} \mathcal{X}^A_{(1,2),\mu} \mathcal{X}^B_{(1,2),\nu} = g_{\mu\nu} + u_{(1,2)\mu} u_{(1,2)\nu}.
\end{eqnarray}
This means that $a_{(1,2)}=n_{(1,2)}|u_1 \cdot u_2|$, and only three of four quantities in (\ref{eq:quantdef}) are independent. Both $a_1$ and $a_2$ can be replaced by the scalar product $u_1\cdot u_2$, which means that the description of the interaction in terms of modified body metric-like objects (\ref{eq:defa}) is equivalent to the description formulated directly through the scalar product $u_1 \cdot u_2$, as in\cite{Andersson,Samuelsson,Jimenez1,Jimenez2}. Therefore, in order to derive the stress-energy tensor, we may rely on results from the literature.

\vskip 2mm
The matter Lagrangian density can be equivalently defined as minus the energy density or as the pressure\cite{Schutz,Seliger}. We will use the definition through the energy density, and consider the matter action of the form
\begin{eqnarray}\label{eq:lagdef}
S_{\textrm{m}} = \int\limits \sqrt{-g} d^4 x \Big[ - \rho_1(n_1,a_1) - \rho_2(n_2,a_2) \Big].
\end{eqnarray}
It is straightforward to generalize this idea to an arbitrary number of interacting fluids, but for the sake of simplicity, we restrict ourselves to the case with only two fluids. The corresponding stress-energy tensor can be derived by varying the matter action with respect to the spacetime metric,
\begin{eqnarray}\label{eq:Tmini}
& & T_{\mu\nu} = - \dfrac{2}{\sqrt{-g}} \dfrac{\delta S_{\textrm{m}}}{\delta g^{\mu\nu}} = T_{1\mu\nu} + T_{2\mu\nu}, \\
& & T_{(1,2)\mu\nu} = -\rho_{(1,2)} g_{\mu\nu} + n_{(1,2)} \dfrac{\partial\rho_{(1,2)}}{\partial n_{(1,2)}} \left(g_{\mu\nu}+u_{(1,2)\mu}u_{(1,2)\nu}\right) \nonumber\\
& & \phantom{T_{(1,2)\mu\nu} = } + a_{(1,2)} \dfrac{\partial\rho_{(1,2)}}{\partial a_{(1,2)}} \left(g_{\mu\nu}-u_{(2,1)\mu}u_{(2,1)\nu} - \dfrac{u_{1\mu}u_{2\nu}+u_{2\mu}u_{1\nu}}{u_1 \cdot u_2} \right). \nonumber
\end{eqnarray}
 The structure of this stress-energy tensor coincides with that derived in\cite{Jimenez1}. It can also be decomposed as
\begin{eqnarray}\label{eq:Tminidecomp}
& & T_{(1,2)\mu\nu} = \mathcal{E}_{(1,2)} u_{(1,2)\mu} u_{(1,2)\nu} + p_{(1,2)\perp} \left(g_{\mu\nu}+u_{(1,2)\mu}u_{(1,2)\nu}\right) \\
& & \phantom{ T_{(1,2)\mu\nu} =} + \left(p_{(1,2)\parallel}-p_{(1,2)\perp}\right) w_{(1,2)\mu}w_{(1,2)\nu} + q_{(1,2)} \left(u_{(1,2)\mu}w_{(1,2)\nu}+u_{(1,2)\mu}w_{(1,2)\nu}\right), \nonumber 
\end{eqnarray}
where $w_{1\mu}$ and $w_{2\nu}$ are normalized space-like $4$-vectors in the direction of the relative motion of fluids in the rest frame of the first and the second fluid, respectively, satisfying
\begin{eqnarray}
w_{(1,2)} \cdot w_{(1,2)} = 1, \quad w_{(1,2)} \cdot u_{(1,2)} = 0, \quad w_{1} \cdot w_{2} = u_{1} \cdot u_{2}, 
\end{eqnarray}
 $\mathcal{E}_{(1,2)}=T_{(1,2)\mu\nu}u_{(1,2)}^{\mu}u_{(1,2)}^{\nu}$ denotes energy densities measured in reference frames of the two fluids, $p_{(1,2)\parallel}$ and $p_{(1,2)\perp}$ are pressures parallel and perpendicular to the direction of relative motion $w_{(1,2)}^{\mu}$, with the anisotropy factor measured by their difference $p_{(1,2)\parallel}-p_{(1,2)\perp}$, and $q_{(1,2)}=-T_{(1,2)\mu\nu}u_{(1,2)}^{\mu}w_{(1,2)}^{\nu}$ denotes energy fluxes of fluids along $w_{(1,2)}^{\mu}$-directions in their own rest frames.

\vskip 2mm
Since $u_1\cdot u_2$ is a scalar, it can be evaluated in the rest frame of either fluid using Riemann normal coordinates. In this way, we can express it through the local relative physical speed of the two fluids $v$ as
\begin{eqnarray}\label{eq:u1u2}
u_1\cdot u_2 = -\dfrac{1}{\sqrt{1-v^2}} = -\gamma_{\textrm{Lor.}}(v),
\end{eqnarray}
with $\gamma_{\textrm{Lor.}}(v)$ denoting the Lorentz relativistic factor. The relative physical speed $v$ then fully describes the interaction of fluids studied in this work.
 By comparing (\ref{eq:Tmini}) with (\ref{eq:Tminidecomp}) we can express quantities parametrizing the latter form of the stress-energy tensor through the relative physical speed as
\begin{eqnarray}\label{eq:eppq}
& & \mathcal{E}_{(1,2)} = \rho_{(1,2)} - \dfrac{v^2}{1-v^2} a_{(1,2)}\dfrac{\partial\rho_{(1,2)}}{\partial a_{(1,2)}}, \\
& & p_{(1,2)\perp} = - \rho_{(1,2)} + n_{(1,2)}\dfrac{\partial\rho_{(1,2)}}{\partial n_{(1,2)}} + a_{(1,2)}\dfrac{\partial\rho_{(1,2)}}{\partial a_{(1,2)}}, \nonumber\\
& & p_{(1,2)\parallel} = - \rho_{(1,2)} + n_{(1,2)}\dfrac{\partial\rho_{(1,2)}}{\partial n_{(1,2)}} + \dfrac{1-2v^2}{1-v^2} a_{(1,2)}\dfrac{\partial\rho_{(1,2)}}{\partial a_{(1,2)}}, \nonumber\\
& & q_{(1,2)} = - \dfrac{v^3}{1-v^2} a_{(1,2)}\dfrac{\partial\rho_{(1,2)}}{\partial a_{(1,2)}}. \nonumber
\end{eqnarray}
Quantities (\ref{eq:quantdef}) entering the action (\ref{eq:lagdef}) can also be evaluated locally in terms of $v$ as
\begin{eqnarray}\label{eq:aann}
n_1=n_0, \quad n_2=\dfrac{n_0}{\gamma_{\textrm{Lor.}}(v)}, \quad a_1=n_0\gamma_{\textrm{Lor.}}(v), \quad a_2=n_0,
\end{eqnarray}
where, in this case, the Riemann normal coordinates have been chosen to be defined in the reference frame of a volume element of the first fluid, in which a volume element of the second fluid moves with speed $v$. In such coordinates, $n_1=n_0$ trivially, and the relation between $n_2$ and $n_0$ is in agreement with relativistic length contraction that rescales volumes of elements of the second fluid with respect to the first fluid by the relativistic factor $\gamma_{\textrm{Lor.}}(v)$. Since $a_2$ measures the particle number density of the second fluid moving with speed $v$ but projected to the frames of the first fluid elements that are static, the relativistic length contraction is canceled with this projection, and $a_2=n_0$. Finally, $a_1$, which measures the density of the first fluid in the rest frame of the second, is affected only by the projection, which yields the factor $\gamma_{\textrm{Lor.}}(v)$. 
 Relations (\ref{eq:aann}) also imply that $a_{(1,2)}=n_{(1,2)}\gamma_{\textrm{Lor.}}(v)$, which is in agreement with the already discussed reduction of the number of independent quantities entering the matter Lagrangian.

\vskip 2mm
The energy density of a fluid with a constant pressure-to-energy density ratio $w$ depends on its particle number density as $\rho\propto n^{w+1}$ at constant entropy per particle. Therefore, in order to compare the two interacting fluids described in this work with non-interacting fluids with constant pressure-to-energy density ratio, we will consider $\rho_1$ and $\rho_2$ of the form
\begin{eqnarray}\label{eq:ro1ro2}
\rho_1(n_1,a_1) = (\rho_1)_0 \left(\dfrac{n_1}{n_0}\right)^{\omega_1+1} \left(\dfrac{a_1}{n_0}\right)^{\lambda_1}, \quad \rho_2(n_2,a_2) = (\rho_2)_0 \left(\dfrac{n_2}{n_0}\right)^{\omega_2+1} \left(\dfrac{a_2}{n_0}\right)^{\lambda_2},
\end{eqnarray}
where $(\rho_1)_0$, $(\rho_2)_0$, $\omega_1$, $\omega_2$, $\lambda_1$ and $\lambda_2$ are constants. The pressure-to-energy density ratios can then be expressed as
\begin{eqnarray}\label{eq:ws}
& & w_{(1,2)\perp} = \dfrac{p_{(1,2)\perp}}{\mathcal{E}_{(1,2)}} = \dfrac{\omega_{(1,2)}+\lambda_{(1,2)}}{1-\dfrac{v^2}{1-v^2}\lambda_{(1,2)}}, \\
& & w_{(1,2)\parallel} = \dfrac{p_{(1,2)\parallel}}{\mathcal{E}_{(1,2)}} = \dfrac{\omega_{(1,2)}+\dfrac{1-2v^2}{1-v^2}\lambda_{(1,2)}}{1-\dfrac{v^2}{1-v^2}\lambda_{(1,2)}}. \nonumber
\end{eqnarray}

\section{Notes on thermodynamics}\label{sec:3}

We have formulated the action (\ref{eq:lagdef}) solely in terms of quantities that depend on the spacetime metric. This approach does not capture thermodynamic properties of fluids, but it suffices for deriving the stress-energy tensor (\ref{eq:Tmini}) and describing gravitational effects, which are the focus of this work. Here we only comment on some key aspects of thermodynamics without providing an exhaustive analysis.

\vskip 2mm
The partial derivative of the energy density with respect to the particle number density can be derived from the first law of thermodynamics,
\begin{eqnarray}\label{eq:ptherm}
n\dfrac{\partial\rho}{\partial n} = \rho + p_{\textrm{th.}},
\end{eqnarray}
with subscript $\textrm{th.}$ indicating that the pressure $p_{\textrm{th.}}$ is defined as a thermodynamic quantity. For a perfect fluid with minimal coupling to gravity and other forms of matter, the thermodynamic pressure $p_{\textrm{th.}}$ defined by this relation coincides with the pressure appearing in the stress-energy tensor, $p_{\textrm{th.}}=p$. This does not hold for the interacting fluids studied in this work, as we can see by comparing (\ref{eq:eppq}) with (\ref{eq:ptherm}),
\begin{eqnarray}
& & p_{(1,2)\perp} = p_{(1,2)\textrm{th.}} + a_{(1,2)} \dfrac{\partial\rho_{(1,2)}}{\partial a_{(1,2)}}, \\
& & p_{(1,2)\parallel} = p_{(1,2)\textrm{th.}} + \left[2-\left(u_1\cdot u_2\right)^2\right] a_{(1,2)} \dfrac{\partial\rho_{(1,2)}}{\partial a_{(1,2)}}, \nonumber\\
& & p_{(1,2)} = \dfrac{2}{3} p_{(1,2)\perp} + \dfrac{1}{3} p_{(1,2)\parallel} = p_{(1,2)\textrm{th.}} + \dfrac{4-\left(u_1\cdot u_2\right)^2}{3} a_{(1,2)} \dfrac{\partial\rho_{(1,2)}}{\partial a_{(1,2)}}, \nonumber
\end{eqnarray}
 where $p_{(1,2)}$ denotes the isotropic average of the anisotropic pressure, and we have used (\ref{eq:u1u2}) as well. This means that thermodynamic effects account for only part of the total pressure; the remainder is due to the interaction. 

\vskip 2mm
The thermodynamic properties of a fluid can be included by adding Lagrange multiplier terms to the action\cite{Brown},
\begin{eqnarray}\label{eq:brown1}
S_{\textrm{m}} = \int\limits d^4 x \Big[ -\sqrt{-g}\rho(n,s) + J^{\mu}\left(\varphi_{,\mu}+s\vartheta_{,\mu}+\beta_{A}\mathcal{X}^{A}_{\phantom{A},\mu}\right) \Big],
\end{eqnarray}
where $s$ is the entropy per particle, $J^{\mu}=\sqrt{-g}j^{\mu}$, with $j^{\mu}$ denoting the particle flux density $j^{\mu}=n u^{\mu}$, and $\varphi$, $\vartheta$ and $\beta_A$ are Lagrange multipliers. Variation of this action with respect to $\varphi$ implies the continuity equation $j^{\mu}_{\phantom{\mu};\mu}=0$, variation with respect to $\vartheta$ leads to the entropy exchange constraint $\left(s j^{\mu}\right)_{;\mu}=0$, and variation with respect to $\beta_{A}$ yields $u^{\mu}\mathcal{X}^{A}_{\phantom{A},\mu}=0$. This means that $\varphi$ is a Lagrange multiplier for conservation of the number of particles, $\vartheta$ enforces conservation of the entropy exchange, and Lagrange multipliers $\beta_{A}$ ensure that the flow lines are the lines of constant Lagrangian coordinates. Thermodynamic variables such as temperature $T$, chemical potential $\mu$ or chemical free energy $f$ are related to the Lagrange multipliers $\varphi$ and $\vartheta$ through relations $T=u^{\mu}\vartheta_{,\mu}$, $\mu=sT+f$, and $f=u^{\mu}\varphi_{,\mu}$, and Lagrange multipliers $\beta_{A}$ are constant along worldlines of volume elements of the fluid.

\vskip 2mm
The most straightforward generalization of the action (\ref{eq:brown1}) to the case of two interacting fluids studied in our work (\ref{eq:lagdef}) can be formulated as
\begin{eqnarray}\label{eq:brown2}
& & S_{\textrm{m}} = \int\limits d^4 x \Big[ -\sqrt{-g}\rho_1(n_1,a_1,s_1) + J_1^{\mu}\left(\varphi_{1,\mu}+s_1\vartheta_{1,\mu}+\beta_{1A}\mathcal{X}^{A}_{1\phantom{|},\mu}\right) \\
& & \phantom{S_{\textrm{m}} = \int\limits d^4 x \Big[}
-\sqrt{-g}\rho_2(n_2,a_2,s_2) + J_2^{\mu}\left(\varphi_{2,\mu}+s_2\vartheta_{2,\mu}+\beta_{2A}\mathcal{X}^{A}_{2\phantom{|},\mu}\right) \Big], \nonumber
\end{eqnarray}
with all quantities defined in the same way as for one fluid. This leads to the same interpretation of the Lagrange multipliers and the thermodynamic quantities as for the single fluid given by the action (\ref{eq:brown1}). In other words, each fluid is associated with its own separate thermodynamic quantities, and the interaction of the two fluids studied in this work modifies only the stress-energy tensor (\ref{eq:Tmini}) and, as we will see in section \ref{sec:5}, the evolution of the relative physical speed $v$. A more complete version of the action (\ref{eq:brown2}) may include additional Lagrange multiplier terms that allow the description of non-adiabatic processes and chemical reactions\cite{Ios}. In this work, we will study purely gravitational effects based on the form of the stress-energy tensor (\ref{eq:Tmini}).

\section{Applications}\label{sec:4}

We have to consider an anisotropic configuration of the two fluids because the interaction has nontrivial effects only when the relative speed is non-zero. This excludes static systems such as non-rotating relativistic stars that are often studied in the literature as objects composed of two fluids, where one component is nuclear matter of a neutron star, and the other is dark matter. The same is also true for rotating stars as long as the two fluids rotate at the same rate, which is the most natural assumption.

\vskip 2mm
Another widely studied field in which matter can be sufficiently well described as a fluid is cosmology. The standard cosmological model --- based on a spatially homogeneous and isotropic Friedmann--Lema\^{i}tre--Robertson--Walker (FLRW) geometry and supplemented by cosmic inflation --- has been remarkably successful in describing a wide range of cosmological observations, such as the large-scale structure of the Universe and the anisotropies of the cosmic microwave background (CMB)\cite{planck1,AlamBOSS,DESI2024,GuthInflation}. Despite this success, several persistent tensions and anomalies suggest that the assumptions underlying the $\Lambda$CDM model require further scrutiny. The most widely discussed example is the Hubble tension: local distance-ladder measurements find a value for the Hubble constant in significant tension with the value inferred from the CMB\cite{RiessSH0ES,planck1,DiValentinoH0}. Another important issue is the so-called $S_8$ tension, in which several weak-lensing analyses have historically preferred lower values for the clustering amplitude than those inferred from the CMB within $\Lambda$CDM\cite{HeymansKiDS,DESY3,planck1}. Additionally, the large-angular-scale CMB sky exhibits several low-multipole features, including alignments and power anomalies\cite{PlanckIso}. These issues do not, by themselves, invalidate the $\Lambda$CDM model. However, they provide strong motivation for studying modifications in which one or more of its assumptions are relaxed. In particular, the assumptions of exact isotropy, comoving matter components, and a purely perfect-fluid stress-energy tensor may be too restrictive. The interaction studied here is thus an example of a possible departure from the $\Lambda$CDM model.

\vskip 2mm
Disregarding small perturbations, the Universe is homogeneous and isotropic on large scales, which requires all its matter components to be comoving. However, at the level of perturbations, the fluids can move relative to each other, and the effects of the interaction appear\cite{Jimenez1,Jimenez2}. We relax the assumption of background isotropy of the Universe by replacing the FLRW spacetime with a Bianchi type-I spacetime in the following section. Moreover, we will restrict ourselves to a special case in which the anisotropy is small, which is strongly preferred by observational restrictions\cite{Akarsu,Aluri} from CMB data \cite{Groen,Pullen} as well as quasar and type Ia supernovae data \cite{Secrest,Sah,Boubel}.

\section{Bianchi type-I spacetime with small anisotropy}\label{sec:5}

We consider an anisotropically expanding Universe with coordinates $(\eta,x,y,z)$, where $\eta$ is conformal time and $x,y,z$ are comoving coordinates. We choose them in such a way that the first fluid is at rest, and the second fluid moves with respect to it in the $x$ direction. Their $4$-velocities are then of the form
\begin{eqnarray}\label{eq:usetup}
u_1^{\mu} = u_1^0 \delta_0^{\mu}, \quad u_2^{\mu} = u_2^0 \left( \delta_0^{\mu} + v_{\textrm{c}} \delta_x^{\mu} \right),
\end{eqnarray}
where $v_{\textrm{c}}=dx/d\eta$ is the $x$-component of the coordinate velocity of the second fluid. The form of the $4$-velocity $u_2^{\mu}$ implies a non-zero component of the stress-energy tensor $T_{0x}$, which requires a non-zero metric component $g_{0x}$. Therefore, we consider the spacetime metric of the form
\begin{eqnarray}\label{eq:bianchi0}
ds^2 = e^{2\alpha} \Big[ -d\eta^2 + 2\gamma d\eta dx + \left(e^{-4\sigma}-\gamma^2\right) dx^2 + e^{2\sigma} \left(dy^2+dz^2\right) \Big],
\end{eqnarray}
where $\alpha$, $\sigma$ and $\gamma$ are functions of the conformal time. The function $\gamma$ parametrizes the $g_{0x}$ component of the metric, but because of it, clocks measuring time $\eta$ are not synchronized, and cannot be synchronized globally by a transformation of the form $\eta_{\textrm{synch.}}=\eta + \delta\eta$ with $\delta\eta$ denoting an appropriate function on spacetime. Such synchronization works locally, $d\eta_{\textrm{synch.}}=d\eta-\gamma dx$, and therefore, we can also locally define the metric induced on hypersurfaces of locally synchronized time as $\gamma_{ij}=g_{ij}-g_{0i}g_{0j}/g_{00}$. For the considered metric (\ref{eq:bianchi0}), we have
\begin{eqnarray}
\gamma_{ij}dx^idx^j = e^{2\alpha} \Big[ e^{-4\sigma} dx^2 + e^{2\sigma} \left( dy^2 + dz^2 \right) \Big].
\end{eqnarray}
The function $\alpha$ then describes the size of space volumes, $V\propto\sqrt{\gamma}=e^{3\alpha}$, which means that $a=e^{\alpha}$ plays the role of the mean scale factor, and $\sigma$ measures the degree of the anisotropy. Since the source of the anisotropy is the motion of the second fluid along the $x$ direction, the metric retains residual isotropy in the $y$-$z$ plane.

\vskip 2mm
Although the metric (\ref{eq:bianchi0}) cannot be globally diagonalized through the time synchronization, a different type of globally well-defined coordinate transformation,
\begin{eqnarray}\label{eq:transf}
\widetilde{\eta} = \int\limits_0^{\eta} \dfrac{d\eta}{\sqrt{1-\gamma^2e^{4\sigma}}}, \quad \widetilde{x} = x + \int\limits_0^{\eta} \dfrac{\gamma e^{4\sigma}}{1-\gamma^2e^{4\sigma}} d\eta, \quad \widetilde{y}=y, \quad \widetilde{z}=z,
\end{eqnarray}
diagonalizes the metric,
\begin{eqnarray}\label{eq:bianchi1}
ds^2 = e^{2\alpha} \Big[ - d\widetilde{\eta}^2 + \left(e^{-4\sigma}-\gamma^2\right) d\widetilde{x}^2 + e^{2\sigma} \left( d\widetilde{y}^2 + d\widetilde{z}^2 \right) \Big].
\end{eqnarray}
This means that the spacetime metric (\ref{eq:bianchi0}) is a special case of a Bianchi type-I metric\cite{Bianchi,Kasner,Ellisb} with residual symmetry in two space dimensions, which, thanks to its simplicity, has been used in various models of the early Universe\cite{Campanelli,Sharif1,Bartolo,Sharif2}. Since $g_{0x}$ can be set to zero by an appropriate choice of coordinates, the same is also true for $T_{0x}$. This corresponds to a procedure described in\cite{Herrera}, where quantities describing two fluids can be redefined in such a way that one fluid can be associated with a $4$-velocity with zero space components, and the other fluid can be associated with a spacelike $4$-velocity-like quantity with zero time component. Instead of employing such an intricate rearrangement, we will work in coordinates in which the spacetime metric is of the form (\ref{eq:bianchi0}) containing the function $\gamma$.

\vskip 2mm
The presence of $\gamma$ in one coordinate system, (\ref{eq:bianchi0}), and its absence in another one, (\ref{eq:bianchi1}), is not contradictory because, as we will see, the function $\gamma$ is not a dynamical variable. In other words, we are considering the ordinary Bianchi type-I metric, which is diagonal in the usual coordinates; the non-diagonal term merely reflects our choice of coordinates adapted to the 4-velocities (\ref{eq:usetup}). The two forms (\ref{eq:bianchi0}) and (\ref{eq:bianchi1}) are physically equivalent. However, using coordinates in which there are non-diagonal components of the stress-energy tensor requires caution. Such non-diagonal components restrict the form of the matter action. Therefore, in what follows, we choose a special form of the functions (\ref{eq:ro1ro2}) entering the matter action (\ref{eq:lagdef}), and then confirm that this choice is consistent.

\vskip 2mm
We will analyze the time evolution of a system with Bianchi type-I spacetime filled with two fluids. The first fluid dominates the total energy density and is not affected by the interaction studied in this work, and the second fluid, affected by the interaction, is a source of a small anisotropy. The matter action can then be chosen as a special form of (\ref{eq:ro1ro2}),
\begin{eqnarray}\label{eq:lagspecial}
S_{\textrm{m}} = \int\limits \sqrt{-g}d^4x \left\{ - \dfrac{\rho_0}{1+\epsilon} \left[ \left(\dfrac{n_1}{n_0}\right)^{w_1+1} + \epsilon \left(\dfrac{n_2}{n_0}\right)^{\omega_2+1}\left(\dfrac{a_2}{n_0}\right)^{\lambda_2} \right] \right\},
\end{eqnarray}
where $\rho_0$ denotes a reference energy density, and $\epsilon$ is a small constant controlling the size of the contribution of the second fluid to the total energy density. The small anisotropy approximation, in which $\epsilon\ll 1$ and $\rho_2/\rho_1$ remains small, is sufficient for determining the criterion for growth or decay of the anisotropy in the late-time limit. The corresponding stress-energy tensor can be written as
\begin{eqnarray}\label{eq:Tmn}
& & T_{\mu\nu} = \rho_1 \left[ (w_1+1)u_{1\mu}u_{1\nu} + w_1 g_{\mu\nu} \right] \\
& & \phantom{T_{\mu\nu} = } + \rho_2 \left[ (\omega_2+1)u_{2\mu}u_{2\nu} + \omega_2 g_{\mu\nu} + \lambda_2 \left( g_{\mu\nu} - u_{1\mu}u_{1\nu} - \dfrac{u_{1\mu}u_{2\nu}+u_{2\mu}u_{1\nu}}{u_1\cdot u_2} \right) \right], \nonumber
\end{eqnarray}
where the pressure-to-energy density ratio of the first fluid is $w_1=\omega_1$, $\rho_1$ and $\rho_2$ are energy densities of the two fluids, and $\rho_1 \gg \rho_2$. Functions parametrizing the spacetime metric (\ref{eq:bianchi0}) and the stress-energy tensor (\ref{eq:Tmn}) can be expanded as 
\begin{eqnarray}\label{eq:Tmnsplits}
& & \alpha = \alpha_{(0)} + \epsilon \alpha_{(1)}, \quad \sigma = \epsilon \sigma_{(1)}, \quad \gamma = \epsilon \gamma_{(1)}, \\
& & \rho_1 = \rho_{1(0)} + \epsilon \rho_{1(1)}, \quad \rho_2 = \epsilon \rho_{2(1)}, \nonumber\\
& & u_{1\mu} = u_{1(0)\mu} + \epsilon u_{1(1)\mu}, \quad u_{2\mu} = u_{2(0)\mu} + \epsilon u_{2(1)\mu}, \nonumber
\end{eqnarray}
and treated perturbatively with respect to $\epsilon$. Contributions that are non-linear in $\epsilon$ are negligible as long as $\epsilon$ is much smaller than $1$, and they will be omitted. Quantities with subscript $(0)$ are then the background or zeroth-order quantities, and subscript $(1)$ denotes the first-order quantities. Normalization of $4$-velocities, $u_1\cdot u_1=u_2\cdot u_2=-1$, implies
\begin{eqnarray}\label{eq:us}
& & u_{1(0)\mu} = - e^{\alpha_{(0)}} \delta_{\mu}^0, \quad u_{1(1)\mu} = - e^{\alpha_{(0)}} \left( \alpha_{(1)} \delta_{\mu}^0 - \gamma_{(1)} \delta_{\mu}^x \right), \\
& & u_{2(0)\mu} = - \dfrac{e^{\alpha_{(0)}}}{\sqrt{1-v_{\textrm{c}}^2}} \left( \delta_{\mu}^0 - v_{\textrm{c}} \delta_{\mu}^x \right), \nonumber
\end{eqnarray}
where the non-zero component $u_{1(1)x}$ is a consequence of non-zero component of the spacetime metric $g_{0x}$. Since $\rho_2$ has no zeroth-order part, $u_{2(1)\mu}$ contributes only at second order and is omitted. For the same reason, it is sufficient to calculate the scalar product $u_1\cdot u_2$ to zeroth-order,
\begin{eqnarray}
u_1 \cdot u_2 = - \dfrac{1}{\sqrt{1-v_{\textrm{c}}^2}} + \mathcal{O}(\epsilon).
\end{eqnarray}
The comparison of this relation with (\ref{eq:u1u2}) then implies that the difference between the coordinate speed $v_{\textrm{c}}$ and the relative physical speed of the fluids $v$ is negligible, and we may set $v_{\textrm{c}}=v$.

\vskip 2mm
Both the Einstein tensor and the stress-energy tensor can be split into the background and the first-order parts,
\begin{eqnarray}
G_{\mu\nu} = G_{(0)\mu\nu} + \epsilon G_{(1)\mu\nu}, \quad T_{\mu\nu} = T_{(0)\mu\nu} + \epsilon T_{(1)\mu\nu},
\end{eqnarray}
and the same is also true for the Einstein field equations,
\begin{eqnarray}\label{eq:efe}
G_{(0)\mu\nu} = 8\pi\kappa T_{(0)\mu\nu}, \quad G_{(1)\mu\nu} = 8\pi\kappa T_{(1)\mu\nu}.
\end{eqnarray}
The zeroth-order equations yield the standard solutions for a flat FLRW spacetime filled with a perfect fluid with constant pressure-to-energy density ratio $w_1$, because the spacetime metric (\ref{eq:bianchi0}) reduces to the FLRW metric and the matter action (\ref{eq:lagspecial}) reduces to the matter action of a single fluid in the zeroth order. The first-order Einstein tensor and stress-energy tensor are written down in Appendix {\ref{app:tensors}}. For $w_1\neq -1/3$, we have
\begin{eqnarray}\label{eq:zeroth}
a = e^{\alpha_{(0)}} = a_0 \left| \dfrac{\eta}{\eta_0} \right|^{\frac{2}{3w_1+1}}, \quad \rho_{1(0)} = \dfrac{1}{8\pi\kappa} \dfrac{12}{\left(3w_1+1\right)^2} \dfrac{e^{-2\alpha_{(0)}}}{\eta^2},
\end{eqnarray}
where $\eta_0$ is a reference conformal time at which $\alpha_{(0)}=\alpha_{(0)0}$ and $a=a_0$. The conformal time is in the interval $\eta\in[0,\infty)$ for $w_1>-1/3$, and in the interval $\eta\in(-\infty,0)$ for $w_1<-1/3$, so that the Universe expands with increasing conformal time. The expansion is accelerated for $w_1<-1/3$, and decelerated for $w_1>-1/3$. The case with $w_1=-1/3$ has to be treated separately, and is treated in Appendix {\ref{app:w1m13}}; in the main text we assume $w_1\neq -1/3$.

\vskip 2mm

We may also check the consistency of the choice of the form of matter action (\ref{eq:lagspecial}) with restrictions given by the presence of non-diagonal components of the stress-energy tensor. Take a more general form of functions (\ref{eq:ro1ro2}) entering the matter Lagrangian (\ref{eq:lagdef}) with $\lambda_1\neq 0$, under the assumption that $\rho_1$ is much larger than $\rho_2$, so that decompositions of quantities into the zeroth and first order in $\epsilon$ (\ref{eq:Tmnsplits}) and (\ref{eq:us}) are valid. Since the non-diagonal component of the zeroth-order Einstein tensor is zero, $G_{(0)0x}=0$, the corresponding component of the stress-energy tensor has to be zero as well,
\begin{eqnarray}
T_{(0)0x} = e^{2\alpha_{(0)}} \rho_{1(0)} \lambda_1 \dfrac{v^3}{1-v^2} = 0. 
\end{eqnarray}
This means that either $v=0$, which is a trivial case with comoving pair of fluids, or $\lambda_1=0$. Hence the choice $\lambda_1=0$ in (\ref{eq:lagspecial}) is not an extra assumption but is enforced by the non-diagonal component.

\vskip 2mm
In order to proceed to the calculation of the first-order quantities, we have to determine the evolution of the relative physical speed. The continuity equation holds for each fluid separately, $j^{\mu}_{(1,2);\mu}=0$, and at zeroth order it yields
\begin{eqnarray}\label{eq:n1n2}
n_1 = b_1 e^{-3\alpha_{(0)}}, \quad n_2 = b_2 e^{-3\alpha_{(0)}} \sqrt{1-v^2},
\end{eqnarray}
where $b_1$ and $b_2$ are constants. Variation of the action (\ref{eq:lagdef}) with respect to the matter fields $\mathcal{X}_1^A$ and $\mathcal{X}_2^A$ yields\cite{Carter,Andersson}
\begin{eqnarray}\label{eq:carter}
j_{(1,2)}^{\mu} \mu_{(1,2)[\mu,\nu]} = 0, \quad \mu_{(1,2)\mu} = \dfrac{\partial\mathcal{L}_{\textrm{m}}}{\partial j_{(1,2)}^{\mu}},
\end{eqnarray}
where $j_{(1,2)}^{\mu}$ denotes particle flux densities of both fluids, and $\mu_{(1,2)\mu}$ denotes the corresponding conjugate momenta defined through the matter Lagrangian density $\mathcal{L}_{\textrm{m}}=-\rho_1-\rho_2$. For the matter action (\ref{eq:lagdef}), we have
\begin{eqnarray}
& & \mu_{(1,2)\mu} = \dfrac{n_0}{n_{(1,2)}} \Bigg[ \left( n_{(1,2)}\dfrac{\partial\rho_{(1,2)}}{\partial n_{(1,2)}} - a_{(2,1)}\dfrac{\partial\rho_{(2,1)}}{\partial a_{(2,1)}} \right) u_{(1,2)\mu} \\
& & \phantom{\mu_{(1,2)\mu} = \dfrac{n_0}{n_{(1,2)}} \Bigg[} - \dfrac{1}{u_1\cdot u_2} \left( a_1\dfrac{\partial\rho_1}{\partial a_1} + a_2\dfrac{\partial\rho_2}{\partial a_2} \right) u_{(2,1)\mu} \Bigg]. \nonumber
\end{eqnarray}
Since all quantities are functions of only the conformal time, equations (\ref{eq:carter}) imply only two independent conditions, $\mu_{1x,\eta}=0$ and $\mu_{2x,\eta}=0$. For the spacetime metric (\ref{eq:bianchi0}) with conventions (\ref{eq:Tmnsplits}) and $4$-velocities (\ref{eq:us}), the $x$-component of conjugate momenta are first-order quantities, $\mu_{(1,2)x}=\epsilon\mu_{(1,2)(1)x}$, of the form
\begin{eqnarray}\label{eq:mu1mu2}
& & \mu_{1(1)x} = \dfrac{n_0}{n_1} e^{\alpha_{(0)}} \left[ \left(\omega_1+1\right) \rho_{1(0)} \gamma_{(1)} + \lambda_2 \rho_{2(1)} v \right], \\
& & \mu_{2(1)x} = \dfrac{n_0}{n_2} e^{\alpha_{(0)}} \left(\omega_2+1\right) \rho_{2(1)} \dfrac{v}{\sqrt{1-v^2}}. \nonumber
\end{eqnarray}
By using both relations in (\ref{eq:n1n2}) and either one of the two equations $\mu_{1(1)x,\eta}=0$ and $\mu_{2(1)x,\eta}=0$, we arrive at
\begin{eqnarray}\label{eq:veta}
v \left(1-v^2\right)^{\frac{\omega_2-1}{2}} = v_0 \left(1-v_0^2\right)^{\frac{\omega_2-1}{2}}\left|\dfrac{a}{a_0}\right|^{3\omega_2+3\lambda_2-1},
\end{eqnarray}
with $v_0$ denoting relative physical speed at the reference time, $v_0=v(\eta_0)$.

\vskip 2mm
This result is in agreement with the differential equation for the relative speed found in \cite{Sandin}, with the only generalization introduced here being the interaction, parametrized by $\lambda_2$. The reason why both equations $\mu_{(1,2)x,\eta}=0$ lead to the same result is that $G^{0}_{(1)x}=0$, which reduces the corresponding component of Einstein field equations to $T^{0}_{(1)x}=0$. This constrains the form of the non-dynamical metric quantity $\gamma_{(1)}$, and as a result of this form of $\gamma_{(1)}$, constant $\mu_{1(1)x}$ implies constant $\mu_{2(1)x}$ and vice versa. In order to avoid cases with absolute values of pressure-to-energy density ratios (\ref{eq:ws}) greater than one, we have to assume $|\omega_2|\leq 1$. Then from the form of (\ref{eq:veta}) we can see that the sign of $\omega_2+\lambda_2-1/3$ determines the late-time behavior of the relative speed. Note also that $\omega_2=-1$ represents a degenerate case in which the time evolution of the relative speed cannot be determined.

\vskip 2mm
There are four independent Einstein field equations for the first-order quantities, namely the $00$, $0x$, $xx$ and $yy$ components ($zz$ coincides with $yy$), see Appendix \ref{app:tensors}. We also have four variables, $\rho_{1(1)}$, $\alpha_{(1)}$, $\gamma_{(1)}$, and $\sigma_{(1)}$, and the remaining quantities are constants $w_1$, $\omega_2$, $\lambda_2$ and the relative speed $v$, which evolves according to (\ref{eq:veta}). The conservation law for the stress-energy tensor can be written in terms of quantities appearing in relation (\ref{eq:carter}) as
\begin{eqnarray}
T_{\mu\nu}^{\phantom{\mu\nu};\mu} = - 2 \left( j_1^{\mu}\mu_{1[\mu,\nu]} + j_2^{\mu}\mu_{2[\mu,\nu]} \right) = 0.
\end{eqnarray}
This is a total conservation law that, owing to the contracted Bianchi identity, contains no information beyond the Einstein field equations. However, the Einstein field equations are the result of variation of the action with respect to the spacetime metric, and such variation does not take into account relative configurations of the two fluids. Therefore, single-fluid models are fully described by Einstein field equations, whereas in two-fluid models, the variation of the matter action with respect to the matter fields $\mathcal{X}_1^A$ and $\mathcal{X}_2^A$ implies new independent equations (\ref{eq:carter}), and with the configuration considered in this paper, only two of them are non-trivial, $\mu_{1(1)x}=\textrm{const.}$ and $\mu_{2(1)x}=\textrm{const}$. Together with (\ref{eq:n1n2}) the conservation law $T_{\mu x}^{\phantom{\mu x};\mu}=0$ implies
\begin{eqnarray}
b_1 \mu_{1(1)x} + b_2 \mu_{2(1)x} = 0,
\end{eqnarray}
where there is an integration constant on the right-hand side which is set to zero by the presence of the non-diagonal component of the stress-energy tensor $T_{(1)0x}$. This leaves us with only one independent equation, which determines the relative physical speed as (\ref{eq:veta}).

\vskip 2mm
The four independent components of the first-order Einstein field equations for four first-order variables can be written directly from relations in Appendix {\ref{app:tensors}}. Here we write down linear combinations of these equations that lead to decoupled equations for each first-order variable. We also introduce a dimensionless conformal time $\widehat{\eta}=\eta/\eta_0$, and rescale the quantities $\alpha_{(1)}$, $\sigma_{(1)}$, $\gamma_{(1)}$ and $\rho_{1(1)}$ using a suitably defined dimensionless constant $K$,
\begin{eqnarray}\label{eq:rescales}
& & \widehat{\alpha} = \dfrac{\alpha_{(1)}}{K}, \quad \widehat{\sigma} = \dfrac{\sigma_{(1)}}{K}, \quad \widehat{\gamma} = \dfrac{\gamma_{(1)}}{K}, \\
& & \widehat{\rho} = \dfrac{\eta_0^2 8\pi\kappa e^{2\alpha_{(0)0}}}{K} \rho_{1(1)}, \quad K = \eta_0^2 8\pi\kappa \rho_{2(1)0} e^{-\left(3\omega_2+3\lambda_2+1\right)\alpha_{(0)0}}, \nonumber
\end{eqnarray}
where $\rho_{2(1)0}$ is $\rho_{2(1)}$ evaluated at $\eta_0$. 

\vskip 2mm
From the $00$ component of Einstein field equations follows
\begin{eqnarray}\label{eq:ro}
& & \widehat{\rho} = \pm \dfrac{12}{3w_1+1} \left(\pm \widehat{\eta}\right)^{-\frac{3w_1+5}{3w_1+1}} \dfrac{d\widehat{\alpha}}{d\widehat{\eta}} - \dfrac{24}{\left(3w_1+1\right)^2} \left(\pm \widehat{\eta}\right)^{-6\frac{w_1+1}{3w_1+1}} \widehat{\alpha} \\
& & \phantom{\widehat{\rho} = } - \left(1-v^2\right)^{\frac{\omega_2+1}{2}} \dfrac{1+\omega_2 v^2}{1-v^2} \left(\pm \widehat{\eta}\right)^{-6\frac{\omega_2+\lambda_2+1}{3w_1+1}}, \nonumber
\end{eqnarray}
the combination $3w_1 G_{00} - G_{ii} = 8\pi\kappa \left(3w_1 T_{00} - T_{ii}\right)$ implies
\begin{eqnarray}\label{eq:alfa}
& & \dfrac{d^2\widehat{\alpha}}{d\widehat{\eta}^2} + \dfrac{2}{\widehat{\eta}} \dfrac{d\widehat{\alpha}}{d\widehat{\eta}} = \dfrac{1}{6} \left(1-v^2\right)^{\frac{\omega_2+1}{2}} \Bigg[ 1-3\omega_2-3\lambda_2 \\
& &
\phantom{\dfrac{d^2\widehat{\alpha}}{d^2\widehat{\eta}} + \dfrac{2}{\widehat{\eta}} \dfrac{d\widehat{\alpha}}{d\widehat{\eta}} = }
+ (3w_1-1) \dfrac{1+\omega_2 v^2}{1-v^2} \Bigg] \left(\pm\widehat{\eta}\right)^{-2\frac{3\omega_2+3\lambda_2+1}{3w_1+1}}, \nonumber
\end{eqnarray}
from $G_{0x} = 8\pi\kappa T_{0x}$ we have
\begin{eqnarray}\label{eq:gama}
\widehat{\gamma} = - \dfrac{\left(3w_1+1\right)^2}{12\left(w_1+1\right)} \left(1-v^2\right)^{\frac{\omega_2+1}{2}} \left( \dfrac{\omega_2+1}{1-v^2} +\lambda_2 \right) v \left(\pm\widehat{\eta}\right)^{6\frac{w_1-\omega_2-\lambda_2}{3w_1+1}},
\end{eqnarray}
and the combination $2G_{xx}-G_{yy}-G_{zz} = 8\pi\kappa \left( 2T_{xx} - T_{yy} - T_{zz} \right)$ yields
\begin{eqnarray}\label{eq:sigma}
\dfrac{d^2\widehat{\sigma}}{d\widehat{\eta}^2} + \dfrac{4}{3w_1+1} \dfrac{1}{\widehat{\eta}} \dfrac{d\widehat{\sigma}}{d\widehat{\eta}} = -\dfrac{1}{3} \left(1-v^2\right)^{\frac{\omega_2+1}{2}} \dfrac{\omega_2+1}{1-v^2} v^2 \left(\pm\widehat{\eta}\right)^{-2\frac{3\omega_2+3\lambda_2+1}{3w_1+1}}.
\end{eqnarray}
We have also used the convention with upper signs corresponding to $w_1>-1/3$, and with lower signs for $w_1<-1/3$.

\vskip 2mm
There are no derivatives of $\gamma_{(1)}$ in equations (\ref{eq:gama}) and (\ref{eq:gamas}). Therefore, as already mentioned, this quantity is not dynamical. By passing to coordinates (\ref{eq:transf}), the metric changes to the form (\ref{eq:bianchi1}), in which $\gamma_{(1)}$ contributes only at the second-order and can be neglected. We can also see that relation (\ref{eq:gama}) is not valid for $w_1=-1$. In this case, terms with $\gamma$ cancel out in the $0x$-component of the Einstein field equations. This implies that $\gamma$ is an arbitrary function of the conformal time, and $\rho_{2}=0$, which means that the second fluid is absent, and we may disregard this trivial case. This is related to the issue of non-diagonal components of the stress-energy tensor imposing constraints on the considered system. The first-order equation $G_{(1)0x}=8\pi\kappa T_{(1)0x}$ now constrains the metric by setting $\gamma_{(1)}$ as (\ref{eq:gama}), and for $w_1=-1$, it also constrains the matter action by setting $\rho_{2}=0$. We can also see that the difference between conformal times $\eta$ and $\widetilde{\eta}$ appearing in (\ref{eq:bianchi0}) and (\ref{eq:bianchi1}) is of the second order. The considered transformation then does not change the form of $\alpha$ and $\sigma$, and therefore, these quantities correctly measure the volume and anisotropy of the Bianchi type-I spacetime.

\vskip 2mm
If $\sigma$ is constant, the anisotropy is not physical, and it disappears in appropriate coordinates. Therefore, in order to describe the physical anisotropy, it is conventional to use the anisotropy parameter defined as $\delta=\dot{\sigma}/\dot{\alpha}$ where the dot denotes differentiation with respect to cosmic time. The anisotropy parameter can be expressed to the first order in $\epsilon$ as
\begin{eqnarray}\label{eq:delta}
\delta = \epsilon K \widehat{\delta}, \quad \widehat{\delta} = \dfrac{3w_1+1}{2} \widehat{\eta} \dfrac{d\widehat{\sigma}}{d\widehat{\eta}},
\end{eqnarray}
where we have also used zeroth-order solutions (\ref{eq:zeroth}).

\vskip 2mm
Evaluation of the anisotropy parameter requires solving the equation for $\sigma$. The solution of the equation (\ref{eq:sigma}) is
\begin{eqnarray}\label{eq:sigmasol}
& & \widehat{\sigma} = \widehat{c}_1 \pm \dfrac{3w_1+1}{3\left(w_1-1\right)} \widehat{c}_2 \left[\left(\pm\widehat{\eta}\right)^{\frac{3\left(w_1-1\right)}{3w_1+1}}-1\right] - \dfrac{1}{3} \int\limits_{\pm 1}^{\widehat{\eta}} d\eta_1 \left(\pm\eta_1\right)^{-\frac{4}{3w_1+1}} \int\limits_{\pm 1}^{\eta_1} d\eta_2 \left(1-v^2\right)^{\frac{\omega_2+1}{2}} \nonumber\\
& & \phantom{\widehat{\sigma} = } \cdot \dfrac{\omega_2+1}{1-v^2} v^2 \left(\pm\eta_2\right)^{-2\frac{3\omega_2+3\lambda_2-1}{3w_1+1}},
\end{eqnarray}
where constants $\widehat{c}_1$ and $\widehat{c}_2$ are given by the initial conditions, $\widehat{\sigma}(\pm 1)=\widehat{c}_1$, and $d\widehat{\sigma}/d\widehat{\eta}(\pm 1)=\widehat{c}_2$. The terms containing these constants form the homogeneous solution. In the special case with $w_1=1$, the homogeneous solution is $\widehat{c}_1 \pm \widehat{c}_2 \ln \left(\pm\widehat{\eta}\right)$ instead of the general form given by (\ref{eq:sigmasol}). The homogeneous solution is not affected by the presence of the second fluid. To isolate the effect of the second fluid, we set $\widehat{c}_1=\widehat{c}_2=0$, and we consider only the particular solution. In this way, we can express $\widehat{\delta}$ defined in (\ref{eq:delta}) as
\begin{eqnarray}
\widehat{\delta} = \mp \dfrac{1}{6}\left(3w_1+1\right)\left(\omega_2+1\right) v_0 \left(1-v_0^2\right)^{\frac{\omega_2-1}{2}} \left(\pm\widehat{\eta}\right)^{\frac{3\left(w_1-1\right)}{3w_1+1}} \int\limits_{\pm 1}^{\widehat{\eta}} d\eta_1 v\left(\eta_1\right), 
\end{eqnarray}
where we have also used (\ref{eq:zeroth}) and (\ref{eq:veta}). 

\vskip 2mm
Using the late-time behavior of the relative speed $v$, given by the sign of $\omega_2+\lambda_2-1/3$ in (\ref{eq:veta}), we can determine the behavior of the anisotropy parameter itself. For $\omega_2+\lambda_2\geq 1/3$, the anisotropy parameter decays in the late-time limit for $w_1<1/3$, tends to a non-zero constant for $w_1=1/3$, and grows for $w_1>1/3$. For $\omega_2+\lambda_2<1/3$, the late-time behavior is instead governed by the sign of $\omega_2+\lambda_2+w_1-2/3$. Consistency of our analysis, however, requires $\omega_2+\lambda_2\geq w_1$. This follows from the requirement that $\rho_2$ does not outgrow $\rho_1$ in the matter action (\ref{eq:lagspecial}), as can be seen from the zeroth-order solutions (\ref{eq:n1n2}) for the quantities (\ref{eq:aann}) entering the action. For $\omega_2+\lambda_2<1/3$, this leaves only $w_1<1/3$, and hence a decaying late-time anisotropy. Both cases can therefore be combined into a single criterion: if either $w_1$ or $\omega_2+\lambda_2$ is smaller than $1/3$, the late-time anisotropy decays.

\vskip 2mm
This generalizes the result of \cite{Sandin}, in which a Bianchi type-I spacetime filled with a pair of non-interacting fluids isotropizes in the late-time limit provided that the pressure-to-energy-density ratio of one of the two fluids is not greater than $1/3$. In the non-interacting limit $\lambda_2=0$, our criterion reproduces this result throughout the region of parameter space consistent with our perturbative scheme. The interaction extends the criterion through the replacement $\omega_2\to\omega_2+\lambda_2$, so that $\lambda_2$ shifts the effective ratio deciding isotropization without altering the threshold value $1/3$ itself. The agreement breaks down only in the marginal cases, in which one of the pressure-to-energy-density ratios equals $1/3$. The reason is that we work to linear order in $\epsilon$, whereas \cite{Sandin} provides a fully non-linear analysis, and the linearization ceases to be self-consistent at late times in the marginal cases. According to (\ref{eq:delta}), a constant anisotropy parameter means that $\sigma_{(1)}$ grows logarithmically, leaving the linear regime. Therefore, we regard the marginal cases as settled in \cite{Sandin} rather than by our linearized treatment.

\begin{figure}[!htb]
\centering
\sbox0{
\includegraphics[scale=0.45]{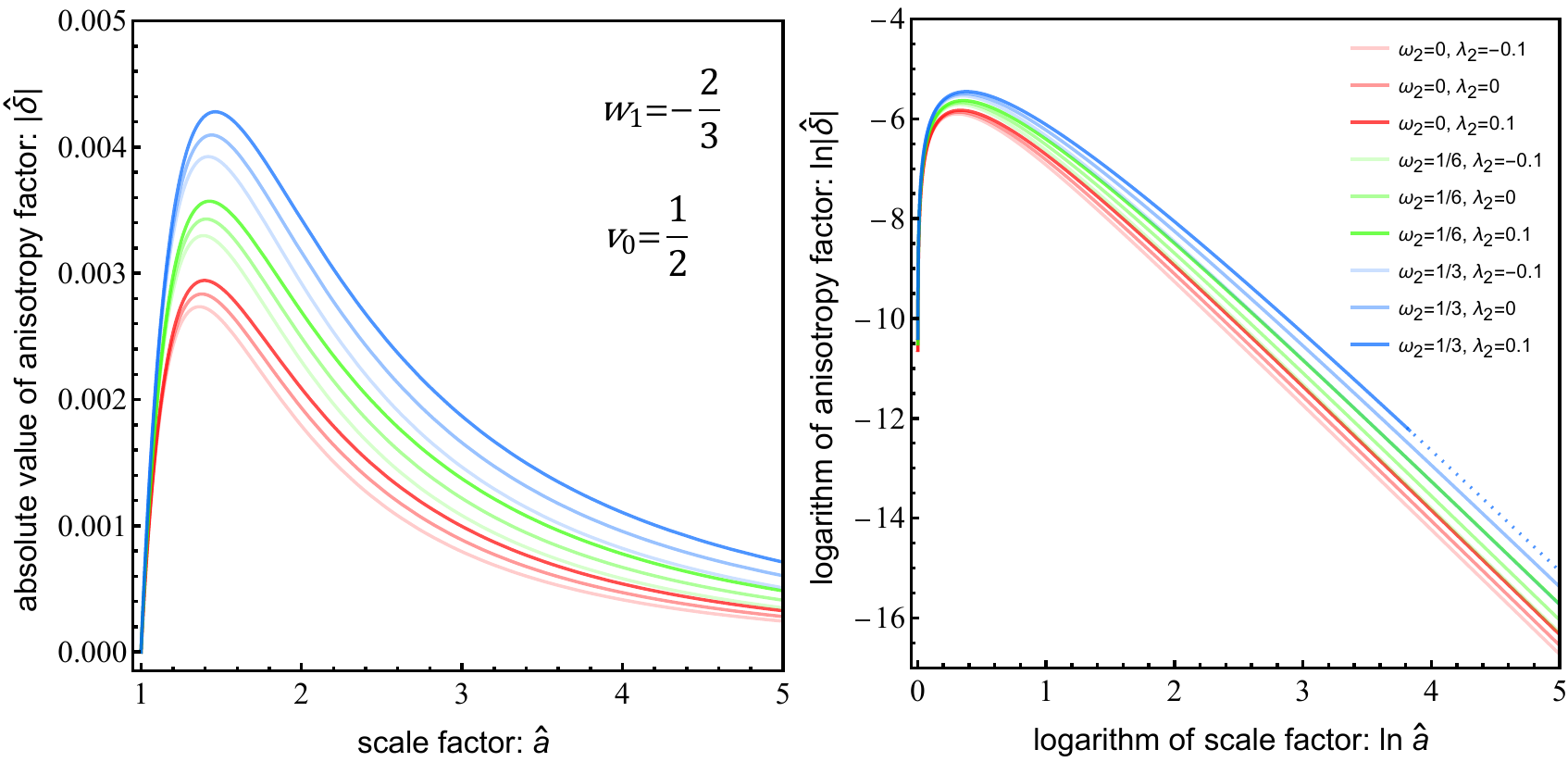}
}
\begin{minipage}{\wd0}
\usebox0
\linespread{1}
\setlength{\abovecaptionskip}{-8pt plus 0pt minus 0pt}
\caption{{\footnotesize 
Dependence of the absolute value of the anisotropy parameter $|\widehat{\delta}|$ on the scale factor $\widehat{a}$ for $w_1=-2/3$, $v_0=1/2$, and various values of $\omega_2$ and $\lambda_2$. In the right panel, we depict natural logarithms of these quantities to capture the late-time asymptotic behavior. Dotted segments indicate the regime in which the absolute value of either parallel or perpendicular pressure-to-energy density ratio (\ref{eq:ws}) exceeds one.}}
\label{fig:wmin1}
\end{minipage}
\end{figure}

\begin{figure}[!htb]
\centering
\sbox0{
\includegraphics[scale=0.45]{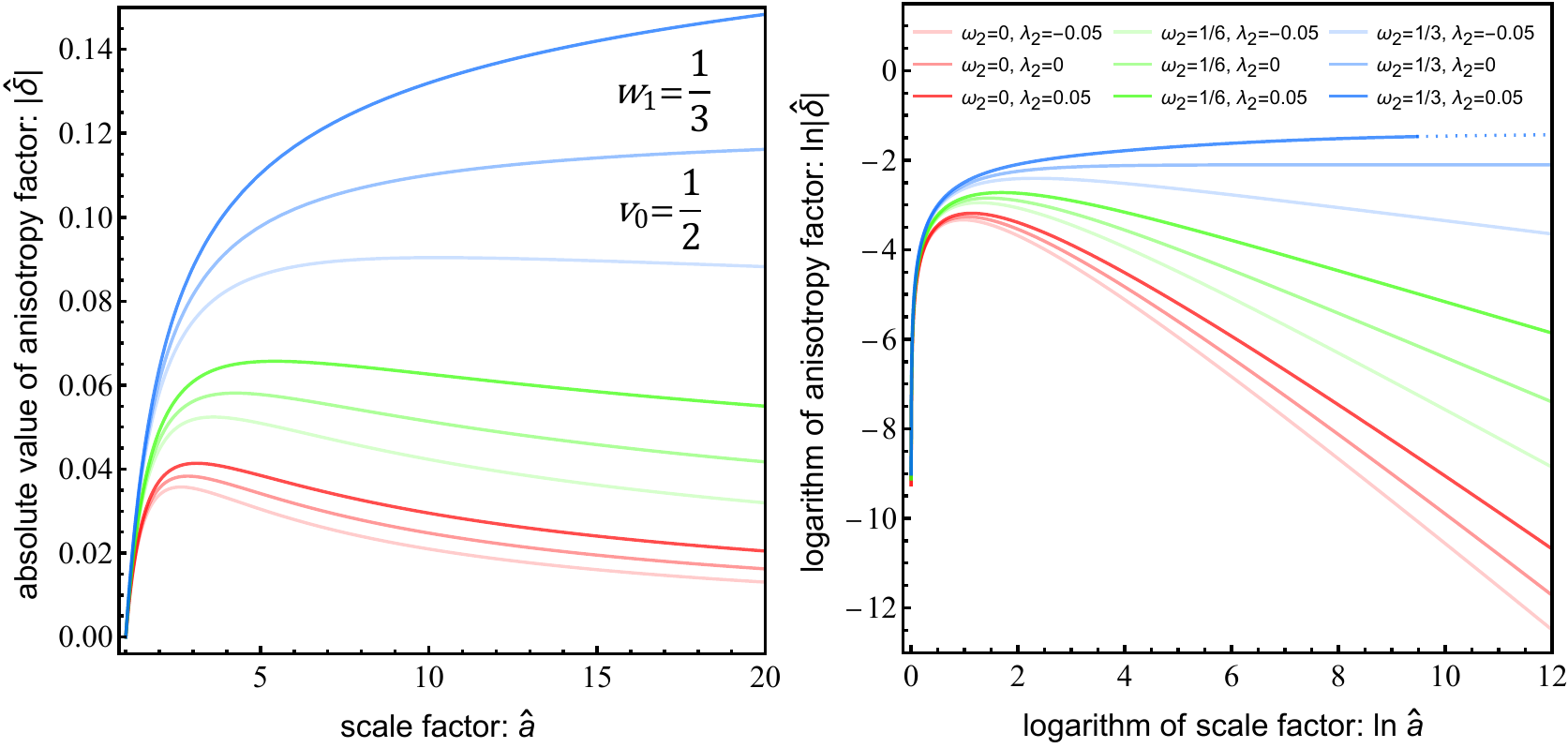}
}
\begin{minipage}{\wd0}
\usebox0
\linespread{1}
\setlength{\abovecaptionskip}{-8pt plus 0pt minus 0pt}
\caption{{\footnotesize Dependence of the absolute value of the anisotropy parameter $|\widehat{\delta}|$ on the scale factor $\widehat{a}$ for $w_1=1/3$, $v_0=1/2$, and various values of $\omega_2$ and $\lambda_2$, with the same conventions as in figure {\ref{fig:wmin1}}.}}
\label{fig:w1third}
\end{minipage}
\end{figure}

\vskip 2mm
In figures \ref{fig:wmin1} and \ref{fig:w1third}, we plot $\widehat{\delta}$ as a function of $\widehat{a}=a/a_{0}=\left(\pm\widehat{\eta}\right)^{2/\left(3w_1+1\right)}$ for various choices of constants $w_1$, $\omega_2$ and $\lambda_2$, and with the time-dependent relative speed given by (\ref{eq:veta}).
Figure {\ref{fig:wmin1}} shows the case with accelerated expansion of the Universe with $w_1=-2/3$. In figure {\ref{fig:w1third}}, $w_1=1/3$ corresponds to a radiation-dominated universe with decelerated expansion. The value of the relative physical speed at the reference time when $\widehat{a}=1$ is chosen as $v_0=1/2$ for both figures. The interaction parameter $\lambda_2$ enters only one exponent in the formula for the relative speed (\ref{eq:veta}), and as we can see, the interaction affects the anisotropy, but does not lead to qualitatively new forms of behavior.

\section{Summary and outlook}\label{sec:6}

Working within the field-theoretical description of fluids, in which the Lagrangian coordinates of the volume elements are treated as fields\cite{Brown}, we have proposed a geometric formulation of a class of fluid interactions previously studied in \cite{Andersson,Samuelsson,Jimenez1,Jimenez2,Ashi}. We have introduced quantities $a_1$ and $a_2$ (\ref{eq:quantdef}) defined as local particle number densities of fluids measured by the spacetime metric induced on hypersurfaces orthogonal not to the fluid's own $4$-velocity but to $4$-velocity of the other fluid (\ref{eq:defa}), i.e., through a modification of the body metric (\ref{eq:def0}). These quantities can then be interpreted as local particle number densities of one fluid measured in reference frames of volume elements of the other fluid. We have obtained the same form of the stress-energy tensor (\ref{eq:Tmini}) as in works where the action is assumed to be directly dependent on the scalar product of particle fluxes\cite{Andersson,Samuelsson} or $4$-velocities\cite{Jimenez1,Jimenez2} of different fluids. Therefore, our work presents an alternative description of the already studied form of interaction of fluids. 

\vskip 2mm
We have studied a system of two fluids filling a Bianchi type-I spacetime. The first fluid dominates the total energy density, and its pressure-to-energy density ratio $w_1$ is constant. The presence of the second fluid introduces a small anisotropy due to its relative velocity with respect to the first fluid, and only the energy density of the second fluid depends on the interaction terms, $\lambda_1=0$. The strength of this interaction is controlled by the constant $\lambda_2$, and another constant $\omega_2$ enters the relation for the parallel and perpendicular pressure-to-energy density ratios of the second fluid (\ref{eq:ws}). The anisotropy decays in the late-time limit if either $w_1$ or $\omega_2+\lambda_2$ is not greater than $1/3$. This generalizes the result of \cite{Sandin}, where the Bianchi type-I Universe is filled with non-interacting fluids, by including the interaction.

\vskip 2mm
The main new result of this paper is that the interaction between fluids shifts one of the quantities determining the criterion for growth or decay of the anisotropy from pressure-to-energy density ratio of the second fluid to $\omega_2+\lambda_2$. This simple shift is in contrast with the considerably more complicated modification of the pressure, which splits into parallel and perpendicular components (\ref{eq:ws}). This means that in the presence of anisotropic stress, the pressure-to-energy density ratios cannot by themselves determine whether the Universe isotropizes. It is to be expected that the same holds for the sound speeds, which should depend not only on the relative speed $v$, but also on the angle between the relative velocity and the direction of propagation. To find the specific form of the sound speeds, one has to consider linearized cosmological perturbations, which are beyond the scope of this work. Another open issue is the stability of the system. The expressions (\ref{eq:ws}) show that, as the relative speed grows, the pressure-to-energy density ratios may exceed one in absolute value, which would violate the dominant energy condition. Since the relative speed grows when $\omega_2+\lambda_2>1/3$, we expect this inequality to be one of the conditions for instability.

\vskip 2mm
Beyond unperturbed Bianchi type-I spacetimes, several extensions are possible. Effects of the interaction on cosmological perturbations on the FLRW background have been studied in\cite{Jimenez1,Jimenez2}. A natural extension is to study perturbations also on anisotropic backgrounds\cite{Gumruk,Pereira}. Other applications may include relativistic stars containing fluids flowing with respect to each other because of rotation\cite{Andersson2}, oscillations\cite{Lindblom}, or the presence of sound waves\cite{Andersson3}; and also further generalizations of fluids, most notably, to relativistic elastic solids\cite{Carter2,Being}.

\section*{Acknowledgements}
We would like to thank to the anonymous referee, whose comments helped us to fix a major error in the original version of our manuscript. The work was supported by grants VEGA 1/0719/23, VEGA 1/0025/23, VEGA 1/0565/25, UK/1345/2026, and Ministry of Education contract No. 0466/2022.

\appendix
\renewcommand{\thesection}{\Alph{section}}

\section{First-order Einstein and stress-energy tensors}\label{app:tensors}

The first-order Einstein tensor is
\begin{eqnarray}
& & G_{(1)00} = 6\alpha_{(0)}'\alpha_{(1)}', \\
& & G_{(1)0x} = -\left(\alpha_{(0)}'^{2}+2 \alpha_{(0)}''\right) \gamma_{(1)}, \nonumber\\
& & G_{(1)xx} = 4\left(\alpha_{(0)}'^{2}+2 \alpha_{(0)}''\right) \sigma_{(1)}-2 \alpha_{(0)}' \alpha_{(1)}'-4 \alpha_{(0)}' \sigma_{(1)}'-2 \alpha_{(1)}''-2 \sigma_{(1)}'', \nonumber\\
& & G_{(1)yy} = -2\left(\alpha_{(0)}'^{ 2}+2 \alpha_{(0)}''\right) \sigma_{(1)}-2 \alpha_{(0)}' \alpha_{(1)}'+2 \alpha_{(0)}' \sigma_{(1)}'-2 \alpha_{(1)}''+\sigma_{(1)}'', \nonumber\\
& & G_{(1)zz} = G_{(1)yy}, \nonumber
\end{eqnarray}
where the prime denotes a derivative with respect to the conformal time $\eta$. For the stress-energy tensor, we have
\begin{eqnarray}
& & T_{(1)00} = e^{2 \alpha_{(0)}}\left[\rho_{1(1)}+2 \rho_{1(0)} \alpha_{(1)}+\rho_{2(1)}\left(\frac{1+\omega_2 v^2}{1-v^2} \right) \right], \\
& & T_{(1)0x} = e^{2\alpha_{(0)}}\left[-\rho_{1(0)}\gamma_{(1)}-\rho_{2(1)}\left(\frac{\omega_2+1}{1-v^2} +\lambda_2 \right) v \right], \nonumber\\
& & T_{(1)xx} = e^{2 \alpha_{(0)}}\left[\rho_{1(1)} w_1+2\rho_{1(0)} w_1\left( \alpha_{(1)}-2 \sigma_{(1)}\right) +\rho_{2(1)}\left(\frac{\omega_2+v^2}{1-v^2} +\lambda_2 \right)\right], \nonumber\\
& & T_{(1)yy} = e^{2 \alpha_{(0)}}\left[\rho_{1(1)} w_1+2\rho_{1(0)} w_1\left( \alpha_{(1)}+ \sigma_{(1)}\right)+\rho_{2(1)}\left(\omega_2 +\lambda_2 \right)\right], \nonumber\\
& & T_{(1)zz} = T_{(1)yy}. \nonumber
\end{eqnarray}

\section{Special case with $w_1=-1/3$}\label{app:w1m13}

For $w_1=-1/3$, the solution for the zeroth-order quantities is
\begin{eqnarray}\label{eq:zeroths}
\alpha_{(0)} = \alpha_{(0)0}\dfrac{\eta}{\eta_0}, \quad \rho_{1(0)} = \dfrac{1}{8\pi\kappa} 3 \left(\dfrac{\alpha_{(0)0}}{\eta_0}\right)^2 e^{-2\alpha_{(0)}}.
\end{eqnarray}
We derive equations for the first-order quantities in the same way as (\ref{eq:ro})-(\ref{eq:sigma}), but instead of the dimensionless conformal time $\widehat{\eta}$ it is more convenient to use $\alpha_{(0)}$ as the time variable,
\begin{eqnarray}\label{eq:ros}
\widetilde{\rho} = 6 e^{-2\alpha_{(0)}} \left( \dfrac{d\widetilde{\alpha}}{d\alpha_{(0)}} - \widetilde{\alpha} \right) - \left(1-v^2\right)^{\frac{\omega_2+1}{2}} \dfrac{1+\omega_2 v^2}{1-v^2} e^{-3\left(\omega_2+\lambda_2+1\right)\alpha_{(0)}},
\end{eqnarray}
\begin{eqnarray}\label{eq:alfas}
\dfrac{d^2\widetilde{\alpha}}{d\alpha_{(0)}^2} = \dfrac{1}{6} \left(1-v^2\right)^{\frac{\omega_2+1}{2}} \left( 1-3\omega_2-3\lambda_2 - 2 \dfrac{1+\omega_2 v^2}{1-v^2} \right) e^{-\left(3\omega_2+3\lambda_2+1\right)\alpha_{(0)}},
\end{eqnarray}
\begin{eqnarray}\label{eq:gamas}
\widetilde{\gamma} = -\dfrac{1}{2} \left(1-v^2\right)^{\frac{\omega_2+1}{2}} \left( \dfrac{\omega_2+1}{1-v^2} +\lambda_2 \right) v e^{-\left(3\omega_2+3\lambda_2+1\right)\alpha_{(0)}},
\end{eqnarray}
\begin{eqnarray}\label{eq:sigmas}
\dfrac{d^2\widetilde{\sigma}}{d\alpha_{(0)}^2} + 2 \dfrac{d\widetilde{\sigma}}{d\alpha_{(0)}} = -\dfrac{1}{3} \left(1-v^2\right)^{\frac{\omega_2+1}{2}} \dfrac{\omega_2+1}{1-v^2} v^2 e^{-\left(3\omega_2+3\lambda_2+1\right)\alpha_{(0)}},
\end{eqnarray}
  with the rescaled quantities 
\begin{eqnarray}
& & \widetilde{\alpha} = \dfrac{\alpha_{(1)}}{L}, \quad \widetilde{\sigma} = \dfrac{\sigma_{(1)}}{L}, \quad \widetilde{\gamma} = \dfrac{\gamma_{(1)}}{L}, \\
& & \widetilde{\rho} = \left(\dfrac{\eta_0}{\alpha_{(0)0}}\right)^2 \dfrac{8\pi\kappa}{L} \rho_{1(1)}, \quad L = \left(\dfrac{\eta_0}{\alpha_{(0)0}}\right)^2 8\pi\kappa \rho_{2(1)0}. \nonumber
\end{eqnarray}
The anisotropy parameter can be calculated as
\begin{eqnarray}
\delta = \epsilon L \widetilde{\delta}, \quad \widetilde{\delta} = \dfrac{d\widetilde{\sigma}}{d\alpha_{(0)}}.
\end{eqnarray}
The solution of (\ref{eq:sigmas}) with initial conditions $\widetilde{\sigma}(1)=\widetilde{c}_1$ and $d\widetilde{\sigma}/d\alpha_{(0)}(1)=\widetilde{c}_2$ is
\begin{eqnarray}
& & \widetilde{\sigma} = \widetilde{c}_1 + \dfrac{1}{2} \widetilde{c}_2 \left[ 1 - e^{2(1-\alpha_{(0)})} \right] \\
& & \phantom{\widetilde{\sigma} = } - \dfrac{1}{3} \int\limits_{1}^{\alpha_{(0)}} d\alpha_{1} e^{-2\alpha_1} \int\limits_{1}^{\alpha_1} d\alpha_2 \left(1-v^2\right)^{\frac{\omega_2+1}{2}} \dfrac{\omega_2+1}{1-v^2} v^2 e^{-\left(3\omega_2+3\lambda_2-1\right)\alpha_2}, \nonumber
\end{eqnarray}
and for the evolution of the anisotropy parameter with $\widetilde{c}_1=\widetilde{c}_2=0$, we have
\begin{eqnarray}
\widetilde{\delta} = -\dfrac{1}{3} \left(\omega_2+1\right) v_0\left(1-v_0^2\right)^{\frac{\omega_2-1}{2}} e^{1-3\omega_2-3\lambda_2} e^{-2\alpha_{(0)}} \int\limits_{1}^{\alpha_{(0)}} d\alpha_1 v(\alpha_1).
\end{eqnarray}

{\setstretch{1.0}

}

\end{document}